Electron Monte Carlo Simulations of Nanoporous Si Thin Films — The Influence of Pore-Edge Charges

Qing Hao[*], Yue Xiao

Department of Aerospace and Mechanical Engineering, University of Arizona

Tucson, AZ 85721-0119

Electronic mail: qinghao@email.arizona.edu

**Abstract**

Electron transport within nanostructures can be important to varied engineering applications, such as thermoelectrics and nanoelectronics. In theoretical studies, electron Monte Carlo simulations are widely used as an alternative approach to solving the electron Boltzmann transport equation, where the energy-dependent electron scattering, exact structure shape, and detailed electric field distribution can be fully incorporated. In this work, such electron Monte Carlo simulations are employed to predict the electrical conductivity of periodic nanoporous Si films that have been widely studied for thermoelectric applications. The focus is on the influence of pore-edge charges on the electron transport. The results are further compared to our previous analytical modeling [Hao *et al.*, *J. Appl. Phys.* 121, 094308 (2017)], where the pore-edge electric field has its own scattering rate to be added to the scattering rates of other mechanisms.

## I. Introduction

At the nanoscale, the electrical properties can be largely affected by the interface or boundary scattering of charge carriers. For thermoelectric (TE) energy conversion, a material should possess a high electrical conductivity $\sigma$, a high Seebeck coefficient $S$, and a low thermal conductivity $k$. The combination of these leads to a high dimensionless figure of merit (ZT), defined as $ZT=S^2\sigma T/k$, where $T$ represents the absolute temperature and the thermal conductivity $k$ can be split into the lattice (phonon) part $k_L$ and the electronic part $k_E$.[1] As one major research direction, various nanostructures are introduced to suppress the phonon transport when the nanostructure size becomes smaller than the mean free paths (MFPs) of majority phonons (see reviews[2-5]). On the other hand, this nanostructure size is still larger than the MFPs of charge carriers to minimize the negative impact on the power factor $S^2\sigma$, leading to a ZT enhanced from that for the bulk counterpart. Examples of this nanostructuring approach include various nanostructured bulk materials,[6-8] nanowires,[9] and nanoporous thin films.[10-12] Although the focus of these studies is normally on the $k_L$ reduction, some changes of electron properties can still be anticipated with the depletion region and the associated potential barrier formed near the surfaces and interfaces, as the result of trapped charges by surface or interfacial defects. In nanoporous Si films, the charge trapping by surface states may deplete the charge carriers and reduce $S^2\sigma$.[11] In other cases, the potential barrier on nanostructured interfaces within a bulk material may possibly benefit $S^2\sigma$ by filtering out low-energy charge carriers, with more gain in $S$ than the reduction in $\sigma$.[13] For nanostructured bulk materials, the energy filtering of charge carriers across embedded interfaces has also been suggested to suppress the transport of minority carriers and thus weaken the detrimental bipolar conduction at elevated temperatures.[14,15]

Accurately predicting the electron transport properties can be challenging for three-dimensional nanostructures, considering the complicated electric field distribution and energy-dependent carrier transport. In a simplified treatment, the scattering rate of a given potential field can be computed and then added to the scattering rates for other mechanisms. The overall scattering rate of charge carriers can then be used in the analytical expressions of electrical properties, as derived from the electron Boltzmann transport equation (BTE).[16] This treatment can be found for ionized impurities within general semiconductors,[17] ionized particles embedded in a film,[18] nanoporous Si films with pore-edge charges,[19] and nanostructured bulk materials with interfacial potential barriers.[20,21] The influence of the local potential field is averaged over the whole solid volume when an effective scattering rate is derived using the Fermi's golden rule. In principle, however, charge carriers are only affected by an electric field within the depletion region. Some errors are thus expected when an effective scattering rate for the whole solid volume is used.

In a more rigorous treatment, numerically solving the electron BTE is not feasible when the distribution function depends on too many parameters, e.g., location, direction, electron band, and energy. In this aspect, electron Monte Carlo (MC) simulations[22-28] should be pursued as an alternative to solving the electron BTE. In these simulations, individual electrons are tracked for their movement and scattering processes. The influence of the energy barrier can be exactly included as the force applied to an electron within the depletion region. When steady states are achieved, the results from these simulations will statistically approach the solution of the BTE. As one big advantage, complicated geometry and detailed energy-dependent electron transport can be fully incorporated.

In this work, electron MC simulations are carried out on two-dimensional periodic nanoporous Si thin films with pore-edge charges. The simulated electrical conductivities for varied pore-edge charges and temperatures are compared to those predicted by an existing analytical model.[19] In general, the two predictions agree with each other, although the analytical model tends to slightly under-predict the electrical conductivity. The electron MC technique presented here can be applied to general porous thin films and two-dimensional materials such as graphene nanomeshes.[29]

**II. Overview of the Electron MC Simulation**

The scheme of electron MC simulations[17,30] is briefly introduced below. More detailed discussions can be found elsewhere.[31] For TE interests, electron MC simulations have been performed on width-modulated nanowires,[22] a one-dimensional chain of grains with a potential barrier at each grain boundary,[23] polycrystalline ceramics,[26] a Si nanostructure with periodic barrier layers.[27] For general applications, such electron MC simulations have also been applied to high-power GaN-based devices[32-34] and heterostructures.[24,25,28]

In electron MC simulations, one period of the nanostructure is created as the computational domain (Fig. 1) with a square-shaped pore located in the center. The perimeter of this square pore should match the perimeter of a circular pore that is assumed in an analytical model.[19] This perimeter match gives very close mean beam lengths (MBLs) for the two porous structures, defined as $MBL = 4V_{Solid}/S$ in radiation studies.[35] For two-dimensional periodic nanoporous films, $V_{Solid}$ is the solid volume within a period and $S$ is the sidewall surface area of a pore. Matched MBLs yield almost identical classical size effects due to boundary scattering on pore edges, regardless of the pore shapes.[36,37]

The computational domain is divided into many spatial bins called "subcells" to statistically determine the local electron states such as the carrier concentration. In simulations, the subcells are usually in rectangular shapes to fit the square-shaped pores and computational domain. To provide more samplings for statistics, each charge carrier is split into multiple identical "partial carriers" particles to make sure that the total number of particles within each subcell is large enough to have statistical significance. Initially, these particles are randomly distributed within the computational domain. On both ends of the computational domain along the x-direction, fixed voltages $\varphi_0$ and $\varphi_L$ are applied. Under the electric field associated with applied $\varphi_0$ and $\varphi_L$, an initial electric field can be computed with COMSOL using the drift-diffusion model and bulk material properties. Driven by this electric field, charge carriers can move with their individual velocities and be scattered during their movement.

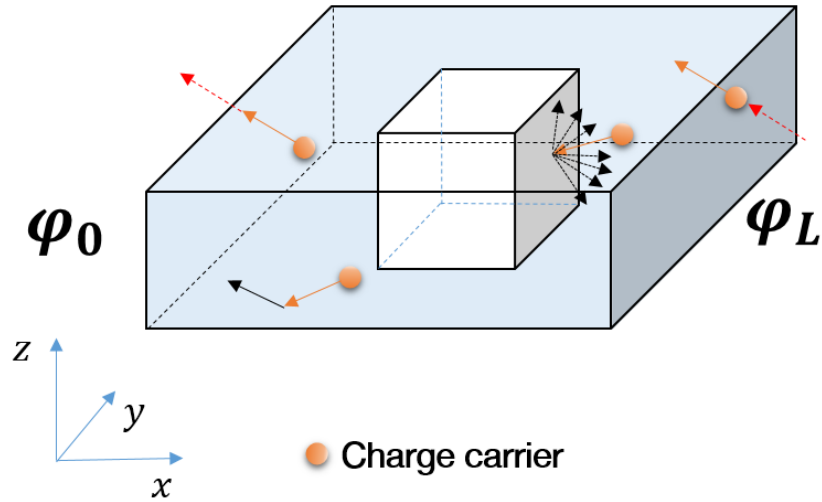

Figure 1. Scheme of the electron MC simulation.

In electron MC simulations, diffusive and elastic reflection is assumed on pore edges. Due to structure symmetry, specular charge-carrier reflection is enforced on two sidewalls (x-z planes) of the computational domain. In addition, the top and bottom film boundaries (x-y planes) also have specular reflection for two-dimensional films, i.e., no influence of the film thickness on the computed electrical properties. If three-dimensional thin films with specific thicknesses are to be considered, a certain degree of diffusive film-surface reflection shall be applied. On both ends of the computational domain (*y-z* planes), a periodic current flux boundary condition is imposed. In this boundary condition, a particle that incidents on one end will be shifted to the same spot on the *y-z* plane of the opposite end to finish its travel for the current time step, keeping its traveling direction and velocity. All employed boundary conditions are illustrated in Fig. 1.

Within the solid volume, charge carriers are mainly scattered by acoustic phonons and ionized impurities for Si. To compute the electrical conductivity at a given temperature $T$, all acoustic phonons are assumed to be in thermal equilibrium with a heat reservoir at $T$. The occurrence of various scattering events is determined by comparing a random number with individual scattering probabilities that are computed from the corresponding scattering rates.[17] To simplify, all internal scattering processes are approximated as elastic scattering. When the electron MC simulation converges, the carrier concentration within each subcell is tallied and fed back into the COMSOL simulation to update the electric field by solving the Poisson equation. In the carrier concentration calculations, the immobile ionized impurities are considered as the background charge density to be added to the local charges contributed by mobile charge carriers. The electron MC simulation is repeated with the updated electric field. The electron MC simulation fully converges when the updated electric field and current density $J_e(y,z)$ counted on both ends are no longer changed. This electric field update has been neglected in some early electron MC studies.[23,26,27] Due to the existence of ballistic electron transport, the nanoscale charge carrier distribution and thus electric field may diverge from those predicted with continuum models developed for bulk materials. This issue is addressed here to ensure accurate description of charge carrier transport within electron MC simulations. When the electron MC simulation converges, the total current $I_{total}$ passing the computational domain should be a constant value. The electrical conductivity $\sigma$ of the simulated structure can then be calculated as

$$\sigma = \frac{I_{total}L}{A(\varphi_L - \varphi_0)}, \qquad (1)$$

where $L$ is the x-direction length and $A$ is the cross-section area of the x-direction ends.

**III. Employed Electronic Band Structure and Scattering Rates of Charge Carriers for Si**

In electron MC simulations, the valley and energy of each partial carrier are randomly set based on the Fermi-Dirac distribution.[17,30] For *n*-type Si, a six-fold degenerate nonparabolic ellipsoidal conduction band is considered here. At an absolute temperature $T$ below $E_g/10k_B$, the contribution of thermally excited minority carriers (i.e., holes for n-type Si) can be simply neglected.[1,38] Here $E_g$ is the band gap and $k_B$ is the Boltzmann constant. For Si, this transition temperature is above 1100 K. Important parameters are given in Table I.

Table I. Parameters used in electron MC simulations.

| Parameter | Symbol | Value |
| --- | --- | --- |
| Electron effective mass, longitudinal | $m^*_{e,l}$ | $0.92\,m_e$ |
| Electron effective mass, transverse | $m^*_{e,t}$ | $0.19\,m_e$ |
| Conduction band degeneracy | $N_{deg}$ | 6 |
| Hole effective mass | $m^*_h$ | $1.2\,m_e$ |
| Bulk modulus | $K$ | $9.8 \times 10^{10}$ N/m² [a] |
| Band gap | $E_g$ | $1.17 - 4.73 \times 10^{-4} T^2/(T+636)$ eV [b] |
| Static permittivity | $\epsilon$ | $20\epsilon_0$ |
| Acoustic phonon deformation potential, electron | $E_{ac}$ | 9 eV |
| Electron concentration | $n$ | $1.0 \times 10^{20}$ cm⁻³ |
| number of charges contributed by each impurity atom | $Z$ | 1 |

(a) Ref [39]

(b) Ref [40]

Other input parameters for electron MC simulations are the energy-dependent scattering rates of individual charge carriers. Within Si, charge carriers are mainly scattered by acoustic phonons and ionized impurities. The scattering rate for the acoustic-phonon scattering of charge carriers is given as[41]

$$\frac{1}{\tau_{AC}} = \frac{\pi E_{ac}^2 k_B T}{N_{deg} K \hbar} D(E), \tag{2}$$

where $k_B$ is the Boltzmann constant, $N_{deg}$ is the band degeneracy, $K$ is bulk modulus of Si, $E_{ac}$ is acoustic phonon deformation potential for electrons, $\hbar$ is the Planck constant divided by $2\pi$. The carrier energy $E$ always refers to the corresponding band extrema and $D(E)$ stands for the electronic density of states (DOS). In Kane's model, $D(E)$ is given by[42,43]

$$D(E) = \frac{\sqrt{2} m_d^{*3/2}}{\pi^2 \hbar^3} \sqrt{E + \alpha E^2}(1 + 2\alpha E). \tag{3}$$

The nonparabolicity $\alpha = 1/E_g$ is used to describe the nonparabolic conduction band. The electronic band structure is described by

$$E(1 + \alpha E) = \frac{\hbar^2}{2}\left(\frac{k_l^2}{m_{e,l}^*} + 2\frac{k_t^2}{m_{e,t}^*}\right), \tag{4}$$

where $k_l$ and $k_t$ are the longitudinal and transverse components of the electron wave vector, respectively. At the band extrema, the electron effective masses along the longitudinal and transverse directions are $m_{e,l}^*$ and $m_{e,t}^*$, respectively. For $D(E)$ and thus $\tau_{AC}$, the electron DOS effective mass is $m_d^* = N_{deg}^{2/3}(m_{e,l}^* m_{e,t}^{*2})^{1/3}$. For semi-classical electron MC simulations, the electron susceptibility effective mass is $m_{\chi,0}^* = 3\left(m_{e,l}^{*-1} + 2m_{e,t}^{*-1}\right)^{-1}$ at the band extrema, i.e., the conduction band minimum. In general, an energy-dependent $m_\chi^*(E) = (1 + 2\alpha E)m_{\chi,0}^*$ is used in the electron MC simulation.[43]

For ionized impurity scattering, the relaxation time $\tau_I$ is given as[17,44]

$$\frac{1}{\tau_I(E)} = \frac{\pi}{8\hbar p^4}\left(\frac{4\pi Z e^2}{\epsilon}\right)^2 D(E) \left\{\ln[1 + (2pr_s)^2] - \frac{(2pr_s)^2}{1+(2pr_s)^2}\right\} N_D, \quad (5)$$

where $Z$ is the number of charges contributed by each ionized impurity atom, $\epsilon$ is the static permittivity, $p$ is the charge-carrier momentum, $N_D$ represents the number density of ionized impurities. With a high $n$-type doping level as $1.0 \times 10^{20}$ cm$^{-3}$, the dopants are completely ionized[45] so that $Z = 1$ can be assumed and $n \approx N_D$ with limited thermally activated carriers. Here $r_s$ stands for the screen length of the Coulomb interaction and is given by

$$r_s^{-2} = \frac{2^{5/2}(m_d^*)^{3/2}e^2\sqrt{k_BT}}{\pi \hbar^3 \epsilon} {}_{1/2}^{0}L_1, \quad (6)$$

where $z$ is reduced charge-carrier energy $z = E/k_BT$, and $f = [1 + \exp(z - \eta)]^{-1}$ is the Fermi-Dirac distribution function with reduced Fermi energy $\eta = E_F/k_BT$. The carrier concentration $n$ can be expressed as

$$n = \frac{(2m_d^* k_B T)^{3/2}}{3\pi^2 \hbar^3} {}_{3/2}^{0}L_0, \quad (7)$$

in which a general integral ${}_m^n L_k$ is defined as

$$_m^n L_k = \int_0^\infty \left(-\frac{\partial f}{\partial z}\right) z^n (z + \beta z^2)^m (1 + 2\beta z)^k dz, \quad (8)$$

with $\beta = k_B T/E_g$. Given a carrier concentration $n$, the Fermi level $E_F$ can be inversely computed by Eq. (7).

## IV. Results and Discussion

In calibrations for bulk Si, the electron MC simulations yield electrical conductivities within 5% divergence from the theoretical predictions at 300–500 K. The same code has also been well calibrated for the high-electric-field electron mobility of bulk GaN or two-dimensional electron gas within GaN-based transistors, where good agreement can be found between the simulations and experimental results.[32-34]

In this work, the electrical conductivity of periodic two-dimensional nanoporous Si films with pore-edge-trapped charges is simulated. The computational domain is a 60 nm × 60 nm square as a single period. Within the computational domain, a square pore with the side length $l$ = 20 nm is located at the center. Across the computational domain, a potential difference of $6 \times 10^{-3}$ V is imposed to create a relatively low electric field (~ $1 \times 10^5$ V/m) for electrical conductivity simulations. Various surface charge density $N_s$ can be added to the pore sidewall (edge), and the corresponding barrier height $U_b$ around a pore can be computed in COMSOL. The final average barrier height $U_b$ around a pore ranges from ~10 to 500 meV in simulated cases. Figure 2 presents the potential distribution with an average $U_b \approx$ 500 meV at 300K.

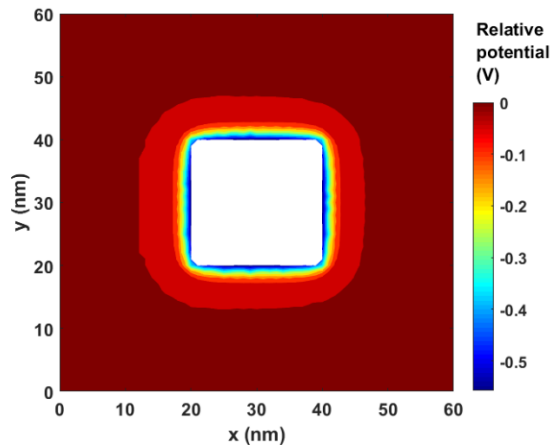

Figure 2. Potential distribution in one period for $U_b \approx$ 500 meV at 300 K.

The simulation results are compared with an analytical model that calculates the electrical properties of a periodic nanoporous film using an effective scattering rate for the pore-edge potential field.[19] For circular pores assumed in the analytical model, the pore perimeter matches the perimeter of the square pore in electron MC simulations. With the same surface charge density $N_s$ on pore edges, the potential barrier height $U_b$ is calculated by

$$U_b = \frac{re^2}{2\epsilon} N_s' \left(\frac{1}{2} \ln \frac{N_s'}{rN_D} - \frac{N_s}{N_s'}\right), \tag{9}$$

where $r$ is the pore radius and $N_s' = rN_D + 2N_s$. With the same surface charge density around the sidewall of the pore, the barrier height $U_b$ calculated with Eq. (9) is generally within 10% variance comparing with the COMSOL results for a square-shape pore. Results from the analytical model and electron MC simulations are compared in Figs. 3a and 3b. Due to enhanced electron scattering, the electrical conductivity $\sigma$ generally decreases at elevated temperatures (Fig. 3a). As anticipated, the electrical conductivity $\sigma$ decreases as the barrier height $U_b$ increases at two representative temperatures (Fig. 3b). In general, the divergence of both predictions is less than 5% and can be very small at zero barrier height.

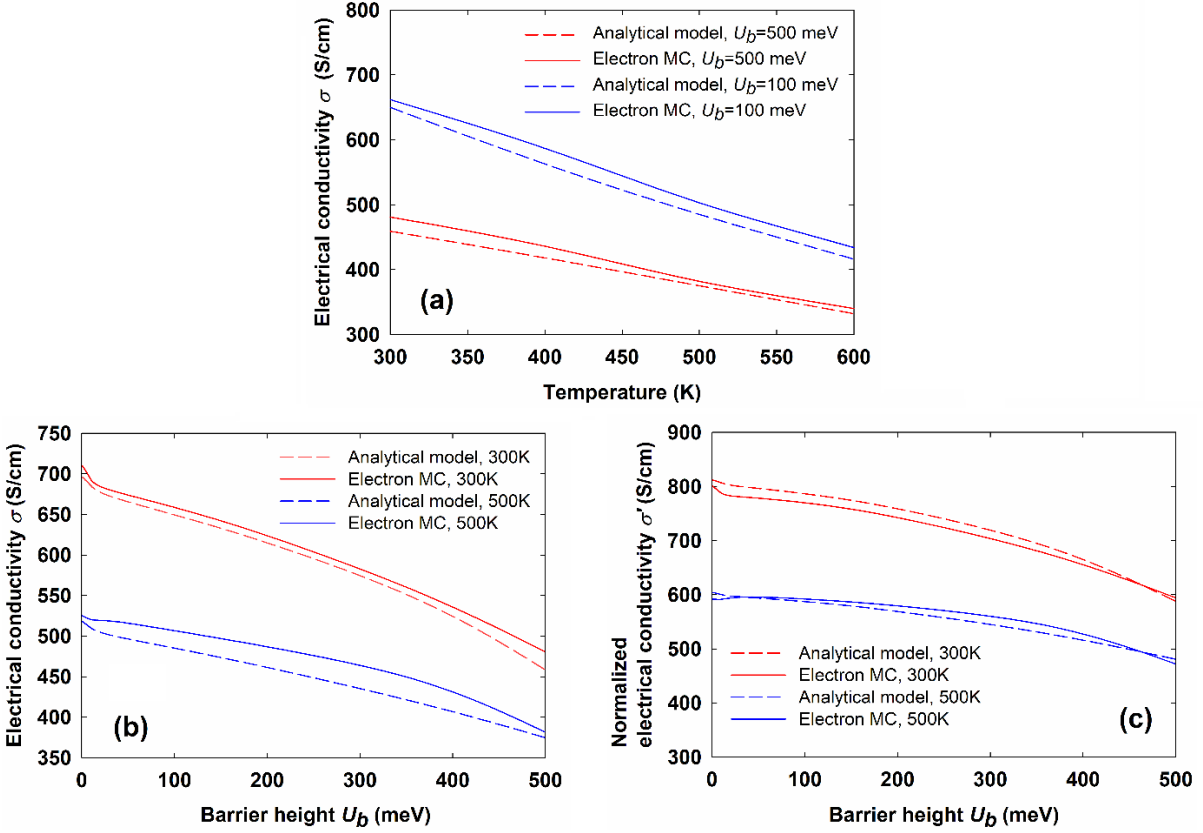

Figure 3. (a) Temperature-dependent $\sigma$ for $U_b$ of 500 meV and 100 meV. (b) The barrier-height-dependent electrical conductivity at 300 K and 500 K. (c) Normalized electrical conductivity $\sigma' = \sigma/F(\varphi)$. The analytical model assumes circular pores, whereas the electron MC simulations assume square pores. All calculations have the same 80 nm pore perimeter with 60 nm pitch of the aligned pores.

Some of the electrical conductivity reduction can be attributed to the reduced number of charge carriers due to pore-edge trapping. Figure 3c presents normalized electrical conductivity with a factor of $F(\varphi) = 1 - \varphi$, where the effective porosity $\varphi$ is determined by the percentage of the area within the depleted region, including the hollow region. This normalized electrical conductivity $\sigma'$ can be viewed as that for a solid thin film with a uniform carrier concentration $n$ and the same pore-edge electric field to scatter charge carriers. In electron MC simulations, some charge carriers can enter the depletion region and get repelled by the electric field. Figure 4 shows the carrier concentration distribution throughout the electron MC computational domain at an averaged $U_b \approx$ 500 meV, where the depletion regions with a width of ~4 nm can be observed around the pore. Unlike a central symmetric circular pore assumed in the analytical model, the depletion width in electron MC simulations varies along the edge of the square pore. The value $1 - \varphi$ is computed from the ratio between the actual number of charge carriers within one period and that computed with a uniform carrier concentration $n$ for the same period of a solid film. For circular pores in the comparison analytical model, the depletion region is 3.2 nm wide.

The normalized $\sigma'$ practically excludes the influence of a reduced number of charge carriers on the electrical conductivity, resulting from pore-edge trapping of nearby charge carriers. With the normalized $\sigma'$, the further reduction of the electrical conductivity can be solely attributed to the electron scattering by the pore-edge electric field. Little divergence can be found between the analytical model and the electron MC simulations in Fig. 3c. Both calculations here show a very similar impact of the pore-edge electric field on the resulting $\sigma'$. For structures with a small neck between adjacent pores and a large $U_b$, the depletion region could overlap and may even completely deplete the material. Such cases are not encountered in the current electron MC simulations.

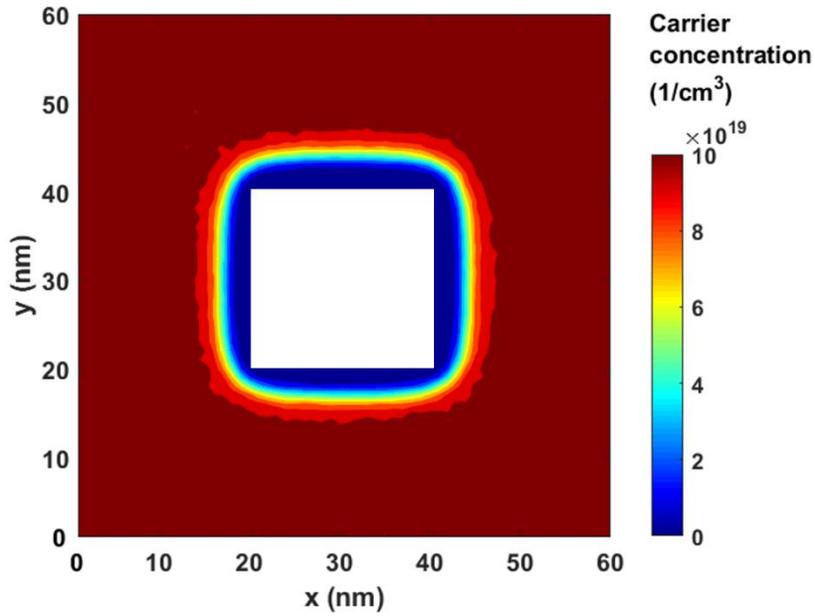

Figure 4. Carrier concentration distribution with an averaged $U_b \approx 500$ meV around the pore.

## V. Summary

An electron MC technique is developed for 2D periodic nanoporous thin films with pore-edge-trapped charges to acquire their electrical conductivities. Compared with an analytical model using an effective scattering rate for the pore-edge electric field, this electron MC simulation considers the detailed interaction between charge carriers and the local electric field around the nanopores. In general, the analytical model and the electron MC simulation render similar results. It can also be observed that the electrical conductivity decreases with increased pore-edge-trapped charges, which can be attributed to the reduced number of charge carriers and enhanced scattering of charge carriers. The MC simulation technique presented in this work can be widely used for other nanostructured materials with the surface- or interface-trapped charges and thus complicated electric fields within the structure.

**Acknowledgements**

This work was supported by the U.S. Air Force Office of Scientific Research [grant number FA9550-16-1-0025] for studies of nanoporous structures, and the National Science Foundation [grant number CBET-1651840] for studies on MC simulations. The authors thank Dr. Ming-Shan Jeng for his help on the electron MC simulations at the beginning. An allocation of computer time from the UA Research Computing High Performance Computing (HPC) and High Throughput Computing (HTC) at the University of Arizona is gratefully acknowledged.